%% file: main.tex
\PassOptionsToPackage{square,comma,numbers,sort&compress,super}{natbib}

\documentclass[reprint,superscriptaddress,amsmath,amssymb,aps,longbibliography,nofootinbib]{revtex4-1}
\usepackage[version=3]{mhchem}
\usepackage[english]{babel}
 \usepackage{amssymb,amsmath,amsthm,mathrsfs,comment,afterpage,accents}
\usepackage{mathtools}
\usepackage{physics} % Adds Dirac braket notation among other helpful bits.
\usepackage{mathrsfs}
\usepackage{amsmath}
\usepackage{booktabs}
\usepackage{gensymb}
\usepackage{graphicx} % Necessary for figures.

\usepackage{xcolor} % Adds more colors for use in text coloring.
\usepackage[colorlinks]{hyperref} % Used for making fancy links.

\newcommand{\eq}[1]{Eq.~\hyperref[eq:#1]{\ref*{eq:#1}}}
\renewcommand{\sec}[1]{\hyperref[sec:#1]{Section~\ref*{sec:#1}}}
\newcommand{\app}[1]{\hyperref[app:#1]{Appendix~\ref*{app:#1}}}
\newcommand{\tab}[1]{\hyperref[tab:#1]{Table~\ref*{tab:#1}}}
\newcommand{\fig}[1]{\hyperref[fig:#1]{Figure~\ref*{fig:#1}}}
\newcommand{\figa}[2]{\hyperref[fig:#1]{Figure~\ref*{fig:#1}#2}}
\newcommand{\figx}[2]{\hyperref[fig:#1]{Figure~\ref*{fig:#1}(#2)}} % chktex 36
\newcommand{\thm}[1]{\hyperref[thm:#1]{Theorem~\ref*{thm:#1}}}
\newcommand{\lem}[1]{\hyperref[lem:#1]{Lemma~\ref*{lem:#1}}}
\newcommand{\cor}[1]{\hyperref[cor:#1]{Corollary~\ref*{cor:#1}}}
\newcommand{\defn}[1]{\hyperref[def:#1]{Definition~\ref*{def:#1}}}
\newcommand{\alg}[1]{\hyperref[alg:#1]{Algorithm~\ref*{alg:#1}}}

\usepackage{dcolumn}% Align table columns on decimal point
\usepackage{bm}% bold maths
\begin{document}
\title{A Non-Orthogonal Variational Quantum Eigensolver}
\author{William J. Huggins}
\email{wjhuggins@gmail.com}
\affiliation{
Department of Chemistry, University of California, Berkeley, CA 94720, USA
}
\affiliation{Berkeley Quantum Information and Computation Center, University of California, Berkeley, CA 94720, USA}
\affiliation{Chemical Sciences Division, Lawrence Berkeley National Laboratory, Berkeley, California 94720, USA}
\author{Joonho Lee}
\altaffiliation{current address: Department of Chemistry, Columbia University, New York, New York 10027, USA}
\affiliation{
Department of Chemistry, University of California, Berkeley, CA 94720, USA
}
\affiliation{Berkeley Quantum Information and Computation Center, University of
  California, Berkeley, CA 94720, USA}
\affiliation{Chemical Sciences Division, Lawrence Berkeley National Laboratory, Berkeley, California 94720, USA}
%\affiliation{Department of Chemistry, Columbia University, New York, New York 10027}

\author{Unpil Baek}
\affiliation{
Department of Physics, University of California, Berkeley, CA 94720, USA
}
\affiliation{Berkeley Quantum Information and Computation Center, University of California, Berkeley, CA 94720, USA}
\affiliation{Chemical Sciences Division, Lawrence Berkeley National Laboratory, Berkeley, California 94720, USA}
\author{Bryan O'Gorman}
\affiliation{Berkeley Quantum Information and Computation Center, University of California, Berkeley, CA 94720, USA}
\affiliation{
Department of Electrical Engineering and Computer Sciences, University of California, Berkeley, CA 94720, USA
}
\affiliation{Chemical Sciences Division, Lawrence Berkeley National Laboratory, Berkeley, California 94720, USA}
\author{K. Birgitta Whaley}
\affiliation{
Department of Chemistry, University of California, Berkeley, CA 94720, USA
}
\affiliation{Berkeley Quantum Information and Computation Center, University of California, Berkeley, CA 94720, USA}
\affiliation{Chemical Sciences Division, Lawrence Berkeley National Laboratory, Berkeley, California 94720, USA}

% \newpage
\begin{abstract}
Variational algorithms for strongly correlated chemical and materials systems
are one of the most promising applications of near-term quantum computers.
We present an extension to the variational quantum eigensolver that approximates
the ground state of a system by solving a generalized eigenvalue problem in a subspace spanned by a collection of parametrized quantum states.
This allows for the
systematic improvement of a logical wavefunction ansatz without a significant increase in
circuit complexity. To minimize the circuit complexity of this approach, we
propose a strategy for efficiently measuring the Hamiltonian and overlap
matrix elements between states parametrized by circuits that commute with the
total particle number operator. We also propose a classical Monte Carlo scheme
to estimate the uncertainty in the ground state energy caused by a finite number
of measurements of the matrix elements. We explain how this Monte Carlo
procedure can be extended to adaptively schedule the
required measurements, reducing the number
of circuit executions necessary for a given accuracy. We apply
these ideas to two model strongly correlated systems, a square configuration of
\ce{H4} and the \(\pi\)-system of Hexatriene (C\(_6\)H\(_8\)).

\end{abstract}

\maketitle

\section{Introduction}

Large, error-corrected quantum computers are expected to
provide powerful new tools for understanding quantum many-body physics. For
example, such devices will be able to efficiently simulate long-time
dynamics~\cite{Lloyd1996-le}, and through phase estimation, measure the energy of a trial
wavefunction while projecting it into the eigenbasis of the
Hamiltonian~\cite{Aspuru-Guzik2005-wz}. Prior to the availability of such devices, it is natural to ask how today's noisy, intermediate-scale quantum
(NISQ) platforms may be used for similar ends. One appealing strategy, 
the variational quantum eigensolver (VQE)~\cite{Peruzzo2014-cv, McClean2016-zj}, uses a potentially noisy
quantum computer as a black box to prepare parametrized wavefunctions and
measure their energy. By optimizing over the wavefunction parameters in a
classical outer loop, one obtains a variational upper bound on the true ground
state energy.

While it is believed that even a noisy, modestly-sized quantum computer can
prepare and measure states that are out of reach for a classical
computer~\cite{Boixo2018-ea}, it will still likely be difficult to take advantage of this fact to surpass the capabilities of
classical variational methods~\cite{Wecker2015-ue, McClean2018-an,
  Kandala2017-tx, Colless2018-ew}. One serious challenge is that noise is particularly damaging for quantum chemical calculations that demand a high degree
of precision~\cite{Kandala2017-tx, Colless2018-ew, Huggins2019-ej}. Recent works
have presented a variety of approaches to overcoming this difficulty, including combining error
detection schemes with postselection~\cite{McArdle2019-pf, Bonet-Monroig2018-pb, Huggins2019-ej}, extrapolating to the zero-noise limit~\cite{Temme2017-yv, Otten2019-td, Kandala2019-hc}, and using additional measurements and post-processing to construct better energy estimators~\cite{McClean2017-uf, Bonet-Monroig2018-pb, McClean2020-em, Sagastizabal2019-sz}. A complementary body of research has focused on developing new variational ans\"atze that use fewer gates and thus offer less opportunity for errors to occur~\cite{Kivlichan2018-is, Lee2019-ne, Grimsley2019-qa}. We shall
present a new approach in this latter direction that allows for a systematic increase in
wavefunction complexity without a growing circuit depth. 

The standard VQE approach uses a
quantum computer to measure the expectation value of the Hamiltonian for some
parametrized wavefunction, \(\ket{\psi(\boldsymbol{\theta})}\), in conjunction with a classical
coprocessor that interprets the measurement outcomes and suggests new values for
the \(\boldsymbol{\theta}\) parameters in order to minimize the energy~\cite{McClean2016-zj}. In
our approach, we define instead a logical ansatz
\begin{equation} \ket{\psi(\mathbf{c}, \boldsymbol{\theta}^{(1)}, \ldots ,
\boldsymbol{\theta}^{(M)})} = \sum_{i=1}^M c_i \ket{\phi_i(\boldsymbol{\theta}^{(i)})},
  \label{eq:psi_definition}
\end{equation} where each \(\ket{\phi_i(\boldsymbol{\theta}^{(i)})}\) is an independently parametrized
wavefunction with a compact quantum circuit description. For brevity, we shall sometimes omit the parameters and refer to these wavefunctions more compactly as \(\ket{\psi}\) and \(\ket{\phi_i}\). Rather than preparing
the state \(\ket{\psi}\) directly on our device and measuring its energy, we use our quantum computer to prepare simpler pairwise superpositions of the
states \(\left\{\ket{\phi_i}\right\}\). We then measure the matrix elements of the
Hamiltonian and overlap matrices,
\begin{gather}
  \label{eq:matrix_elements}
H_{ij} = \mel{\phi_i}{\hat{H}}{\phi_j}, \\
S_{ij} = \braket{\phi_i}{\phi_j}. \nonumber
\end{gather} 
This allows us to classically solve a generalized eigenvalue problem,
\begin{equation} H\mathbf{c} = E S \mathbf{c},
    \label{eq:generalized_eigenvalue_problem_1}
\end{equation} thereby finding the optimal \(\mathbf{c}\) parameters and minimizing the energy in the subspace spanned by the
set of states \(\{\ket{\phi_i}\}\). The \(\boldsymbol{\theta}^{(i)}\) values that
parametrize each basis function \({\ket{\phi_i(\boldsymbol{\theta}^{(i)})}}\) can then
be optimized by a classical outer loop to lower the energy further, solving a new generalized eigenvalue problem at each step.

Our approach shares certain features with a variety of recent proposals for quantum
algorithms that involve solving generalized eigenvalue
problems~\cite{McClean2017-uf, Motta2020-in, Parrish2019-ev, Kyriienko2020-ez, Parrish2019-hf, Stair2019-sc}. However, our approach also differs from these works in some key respects. Most importantly, we
make no assumptions about the form of the component wavefunctions
\(\ket{\phi_i}\), other than that they have efficient quantum circuit
implementations. In the context of quantum algorithms, prior work has assumed that these wavefunctions are generated by excitations from a fixed
reference state~\cite{McClean2017-uf}, by real or imaginary time
evolution~\cite{Kyriienko2020-ez, Motta2020-in, Parrish2019-hf, Stair2019-sc}, or by the simultaneous rotation of a set
of orthogonal reference wavefunctions~\cite{Parrish2019-ev}. Two of these works in particular, Refs.~\citenum{Parrish2019-hf} and \citenum{Stair2019-sc}, were released contemporaneously with our own and provide an interesting contrast to our approach. Specifically, they require the same offdiagonal matrix element measurements used in this work but construct the non-orthogonal basis function by real-time propagation of trial wavefunctions rather than the variational approach we take here.

In the context of classical simulations, multireference methods which make use of a superposition of configurations have a long and storied history~\cite{Condon1951-sb, Jeziorski1981-pb, Werner1982-mq, Koch1993-mo, Malmqvist1986-nz, Thom2009-fp, Jimenez-Hoyos2013-wb, McClean2015-es, Sundstrom2014-rb, Landinez_Borda2019-bj}. Most directly similar to this work are those which demand each of the \(\ket{\phi_i}\)
wavefunctions to be a Slater determinant (not necessarily in the same single
particle basis)~\cite{Koch1993-mo}. This basic direction has been elaborated upon under a variety of names, including the non-orthogonal configuration interation (NOCI) method~\cite{Malmqvist1986-nz, Thom2009-fp, Sundstrom2014-rb}, the non-orthogonal Multicomponent Adaptive Greedy Iterative  Compression (NOMAGIC) algorithm~\cite{McClean2015-es}, and the non-orthogonal multi-Slater determinant (NOMSD) expansion approach ~\cite{Jimenez-Hoyos2013-wb, Landinez_Borda2019-bj}, among others. The restriction to Slater determinants allows for the efficient classical evaluation of the required matrix elements while the relaxation of the requirement that the determinants be orthogonal to one another allows for more flexible and accurate wavefunctions when compared to orthogonal CI expansions with the same number of determinants. 

The difference between these various approaches mainly lies in the way in which they obtain a set of non-orthogonal determinants. For example, NOCI separately optimizes individual
determinants by finding a collection of different solutions to the Hartree-Fock equations before performing a single diagonalization of the Hamiltonian matrix~\cite{Malmqvist1986-nz, Thom2009-fp, Sundstrom2014-rb}. Other approaches more closely parallel the one we take here, iteratively adding new states and variationally optimizing their parameters~\cite{Koch1993-mo, Jimenez-Hoyos2013-wb, Landinez_Borda2019-bj}. We do not exhaustively review the classical literature, but note that the variational approach has been found to be prone to optimization challenges and that a number of the methods we cite arise out of attempts to ameliorate this difficulty~\cite{McClean2015-es, Jimenez-Hoyos2013-wb, Landinez_Borda2019-bj}.

By taking the basis functions \(\ket{\phi_i}\) to be independently parametrized quantum circuits rather than single Slater determinants, we obtain an extremely flexible
form for our logical ansatz, \(\ket{\psi} = \sum c_i \ket{\phi_i}\). For a wide variety of ansatz circuits, we shall show that the required matrix
element measurements between any \(\ket{\phi_i}\) and \(\ket{\phi_j}\) pair can be
implemented efficiently using a number of quantum gates that is equal to the sum of the
gates required to prepare \(\ket{\phi_i}\) and \(\ket{\phi_j}\), plus a small
factor that scales linearly with the system size.
Notably, the quantum volume required is independent of the number of wavefunctions in the logical ansatz, making it
possible to systematically add flexibility to \(\ket{\psi}\) without increasing
the required gate fidelity or coherence times of the quantum hardware.

This flexibility, however, comes at the cost of demanding more
matrix element measurements. To ameliorate this cost we propose using a Monte
Carlo technique to estimate the uncertainty in the ground state energy and to adaptively allocate our measurements of the
matrix elements. Essentially, this scheme involves sampling from the distributions representing the
uncertainty in the estimates of the Hamiltonian and overlap matrices and solving a small
generalized eigenvalue problem for each sampled matrix pair. We then characterize the resulting distribution of ground state energy values by a sample variance. We suggest a heuristic that repeatedly determines
which measurement to perform by calculating the sensitivity of this sample
variance to additional measurements of each of the matrix elements.

We apply these ideas to two model chemical systems, a square
configuration of H$_4$ and the \(\pi\)-system of hexatriene (C$_6$H$_8$), which exhibit mixed
strong correlation and dynamical correlation effects. In terms of strong
correlation, we shall focus on a pair of strongly entangled electrons. Specifically,
there can be two exactly degenerate determinants for certain geometries of these
systems while the rest of the electrons contribute to dynamical correlation. We
present two types of numerical experiments. In the first, we explore how well
the ground state of these systems can be represented by an NOVQE logical ansatz,
varying both the complexity of the constituent basis functions and the size of
the subspace. In the second, we take a fixed set of basis wavefunctions and
compare our adaptive protocol for scheduling measurements with a simpler alternative.

\section{Theory}

\subsection{Matrix Element Measurement}
The off-diagonal matrix elements of the Hamiltonian, \(H_{ij} =
\mel{\phi_i}{\hat{H}}{\phi_j}\), do not correspond to physical observables and
therefore cannot be measured directly in the usual manner. Nevertheless, it
is possible to construct circuits that allow us to estimate them, for example,
by using the Hadamard test~\cite{Yu_Kitaev1995-wa}. In
this section we present a simple strategy for measuring these
matrix elements. We combine ideas from recent proposals for measuring
off-diagonal matrix elements that appear in other contexts~\cite{Schuld2019-rb,
  Izmaylov2020-zt, Kyriienko2020-ez} with a trick inspired by the literature on the impossibility
of black box protocols for adding controls to arbitrary
unitaries~\cite{Araujo2014-vq}. Our strategy offers several benefits over a naive
application of the Hadamard test. Namely, it doesn't require implementing
controlled versions of the ansatz preparation circuits, and it enables the
simultaneous measurement of matrix elements of multiple commuting observables
while also yielding information about the overlap matrix elements, \(S_{ij} =
\braket{\phi_i}{\phi_j}\).

For simplicity, we will describe below the
case where \(\hat{H}\) is a sum of commuting operators which can easily be
simultaneously measured. In the more general case, the usual Hamiltonian
averaging approach of grouping the terms into multiple sets that are each
simultaneously measurable and measuring the sets separately can be applied
without modification~\cite{McClean2016-zj, Kandala2017-tx, Rubin2018-ir}.

We begin by preparing the state
\begin{equation} \ket{+_{ij}} := \frac{1}{\sqrt{2}}(\ket{\phi_i}\ket{0} +
\ket{\phi_j}\ket{1}),
\end{equation}
where the second register is an ancilla qubit. This task can be accomplished by using controlled versions of the unitaries \(\hat{U}_i\) and
\(\hat{U}_j\) that prepare \(\ket{\phi_i}\) and \(\ket{\phi_j}\) from a fixed
reference state. Given some quantum circuit that implements the unitaries
\(\hat{U}_i\) and \(\hat{U}_j\), it is possible to construct circuits that implement the controlled
version of \(\hat{U}_i\) and \(\hat{U}_j\), by replacing each gate in the
original circuits with its controlled form. Even setting aside the difficulty of
compiling such a circuit on a physical device with limited connectivity, the cost of implementing such a circuit on a near-term device
(quantified by counting the number of two-qubit gates) will be
substantially increased. For example, it is known that the decomposition of the
Toffoli gate (the controlled-controlled-NOT gate) into a collection of single qubit and
CNOT gates requires the use of six CNOT gates~\cite{Shende2008-sv}. Given the
limited coherence times and two-qubit gate fidelities of near-term hardware, we must ask if there are alternatives for implementing controlled
versions of \(\hat{U}_i\) and \(\hat{U}_j\).

An ideal protocol might allow us to implement a controlled version of an arbitrary
\(\hat{U}\) using a single execution of the original, unmodified circuit that implements \(\hat{U}\). Unfortunately, a single use of oracle
(blackbox) access to a general \(\hat{U}\) is insufficient for implementing a
controlled version of \(\hat{U}\) in the quantum circuit
model~\cite{Araujo2014-vq}. However, if \(\hat{U}_i\) and \(\hat{U}_j\) preserve fermionic (or
bosonic) excitation number and act trivially on the vacuum state, then we can circumvent this no-go result. We now show how this can be accomplished in the construction of a controlled unitary operator,
\begin{equation}
  \hat{U}_i, \hat{U}_j \rightarrow \hat{U}_i \otimes
  \ketbra{0}{0} + \hat{U}_j \otimes \ketbra{1}{1}.
\end{equation}

We begin with a generic input state
\(\ket{\psi_0}\ket{0} + \ket{\psi_1}\ket{1}\), subject to the restriction that
\(\ket{\psi_0}\) and \(\ket{\psi_1}\) are both states that are orthogonal to the
state with zero particles, \(\ket{\mathrm{vac}}\).
\begin{enumerate}
\item First, we adjoin an ancilla system register in the vacuum state to obtain

\(\ket{\psi_0} \otimes \ket{\mathrm{vac}} \otimes \ket{0} + \ket{\psi_1} \otimes
\ket{\mathrm{vac}} \otimes \ket{1}\).

\item Treating the final qubit as the control, we apply a controlled-SWAP operation between the two system registers, resulting in
 
\(\ket{\psi_0} \otimes \ket{\mathrm{vac}} \otimes \ket{0} + \ket{\mathrm{vac}} \otimes
\ket{\psi_1} \otimes \ket{1}\).

\item Next, we execute the unmodified circuit for \(\hat{U}_i\) on the first system
register, while doing the same with \(\hat{U}_j\) on the second
system register, yielding

\(\hat{U}_i \ket{\psi_0} \otimes \ket{\mathrm{vac}} \otimes \ket{0} + \ket{\mathrm{vac}} \otimes
\hat{U}_j \ket{\psi_1} \otimes \ket{1}\).
  
\item We follow this with a second controlled-SWAP operation to produce the state,

\(\hat{U}_i \ket{\psi_0} \otimes \ket{\mathrm{vac}} \otimes \ket{0} + \hat{U}_j
\ket{\psi_1} \otimes \ket{\mathrm{vac}} \otimes \ket{1}\).

\item Finally, we discard the now unentangled second system register to show
  completion of the action
of the controlled unitary gate and obtain the desired result,

\(\hat{U}_i \ket{\psi_0} \otimes \ket{0} + \hat{U}_j \ket{\psi_1} \otimes
\ket{1}\).
\end{enumerate}

For our purposes, we can take \(\ket{\psi_0}\) and \(\ket{\psi_1}\) to be the
same fixed reference state, usually a Hartree-Fock state \(\ket{\psi_{\mathrm{HF}}}\).
Then \(\ket{\phi_i} = \hat{U}_i\ket{\psi_{\mathrm{HF}}}\) and \(\ket{\phi_j} =
\hat{U}_j\ket{\psi_{\mathrm{HF}}}\) and we see that with the last step we have successfully prepared the desired state, \(\ket{+_{ij}} := \frac{1}{\sqrt{2}}(\ket{\phi_i}\ket{0} +
\ket{\phi_j}\ket{1})\). We then apply a Hadamard gate on the ancilla qubit and perform a
\(\hat{Z}\) measurement. It is easy to see that the expectation value of
\(\hat{Z}\) for the ancilla qubit will be
\(\langle\hat{Z}_{\mathrm{anc}}\rangle = \mathrm{Re}\braket{\phi_i}{\phi_j}\).
Furthermore, the post-measurement state of the system register is either
\begin{equation}
  \frac{\ket{\phi_i} + \ket{\phi_j}}{\sqrt{2 + 2 \mathrm{Re}
      \braket{\phi_i}{\phi_j}}},
\end{equation}
if the ancilla qubit was found to be in the \(+1\) eigenstate, or
\begin{equation}
  \frac{\ket{\phi_i} - \ket{\phi_j}}{\sqrt{2 - 2 \mathrm{Re}
      \braket{\phi_i}{\phi_j}}},
\end{equation}
if the measurement outcome was \(-1\). These outcomes occur 
with probabilities
\(\frac{1 + \mathrm{Re}\braket{\phi_i}{\phi_j}}{2}\)
and
\(\frac{1 - \mathrm{Re}\braket{\phi_i}{\phi_j}}{2}\) respectively.

In both cases, we proceed to measure the Hamiltonian \(\hat{H}\) on the system register.
Depending on the result of the ancilla qubit measurement, the resulting
expectation values will be either
\begin{equation}
  \langle\hat{H}\rangle^{(+1)} =
  \frac{\langle\hat{H}\rangle_i +
    \langle\hat{H}\rangle_j +
    2 \mathrm{Re}\mel{\phi_i}{\hat{H}}{\phi_j}}{
    2 + 2 \mathrm{Re}\braket{\phi_i}{\phi_j}},
    \label{eq:H_plus_plus_one_ev}
\end{equation}
  or
\begin{equation}
  \langle\hat{H}\rangle^{(-1)} =
  \frac{\langle\hat{H}\rangle_i +
    \langle\hat{H}\rangle_j - 2 \mathrm{Re}\mel{\phi_i}{\hat{H}}{\phi_j}}{ 2 - 2 \mathrm{Re}
    \braket{\phi_i}{\phi_j}}.
    \label{eq:H_plus_minus_one_ev}
\end{equation}
Now we consider the expectation value of the operator \(\hat{H} \hat{Z}_{\mathrm{anc}}\).
By multiplying each of the conditional expectation values of \(\hat{H}\) by the
corresponding eigenvalue of \(\hat{Z}_{\mathrm{anc}}\) and taking the appropriate
weighted average, we find that
\begin{equation}
  \langle \hat{H} \hat{Z}_{\mathrm{anc}} \rangle =
  \mathrm{Re}\mel{\phi_i}{\hat{H}}{\phi_j}.
  \label{eq:HZ}
\end{equation}
Furthermore, if \(\hat{H}\) is a sum of Pauli operators, then the usual Hamiltonian averaging approach and upper bounds on the variance of a VQE
observable apply to \eq{HZ}~\cite{Rubin2018-ir}. Therefore, by
repeated measurement we can estimate \(\mathrm{Re}\mel{\phi_i}{\hat{H}}{\phi_j}\) to
a fixed precision \(\epsilon\) using approximately the same number of
measurements that we would need to measure a diagonal matrix element to the same
accuracy. A similar approach allows us to estimate \(\mathrm{Im}\braket{\phi_i}{\phi_j}\)
and \(\mathrm{Im}\mel{\phi_i}{\hat{H}}{\phi_j}\) by starting with the state
\(\frac{1}{\sqrt{2}}(\ket{0}\ket{\phi_i} + i \ket{1}\ket{\phi_j})\).

Consider an ansatz $\ket{\theta} = U(\theta)\ket{\psi_0}$ on $N$ qubits such that the size and depth of the circuit for $U$ is independent of $\theta$; this is typical of VQE ans\"atze, but the following can be generalized easily when it is not the case.
Suppose also that we have a protocol for measuring the Hamiltonian $H$ on the $N$-qubit register.
What are the additional resources required to implement NOVQE?\@
First, we require $2N$ qubits and at least one ancilla.
The variational unitaries $U_i$ and $U_j$ can be applied in parallel, doubling the size of the circuit but not the depth.
The measurement protocol for $H$ can be applied without modification to the first register.
For the two controlled swaps, there is a space-time tradeoff.
First, consider the case without geometric constranits.
Each controlled swap of the registers can be implemented using the single ancilla and $N$ 3-qubit CSWAP gates in series on pairs of the corresponding qubits from the two registers, adding $2N \tau_{\mathrm{CSWAP}}$ to the depth, where $\tau_{\mathrm{CSWAP}}$ is the effective depth of the CSWAP gate.
Alternatively, we can use $N$ ancilla and in $\left\lceil\log_2 N\right\rceil$ depth produce a cat state.
Then the $N$ CSWAPS can be done in parallel, adding only $2\tau_{\mathrm{CSWAP}}$ to the depth.

Suppose now that we are restricted, e.g., to some subgraph of a 2D square grid, and that $U(\theta)$ can be implemented only using gates on linearly adjacent qubits.
Then we can place the computational registers on adjacent rows and the ancilla at the end of one.
Now, in addition to the CSWAP gates, we must use $N$ 2-qubit SWAP gates to move the ancilla through the line, so that the contribution to the depth is now $2 N (\tau_{\mathrm{CSWAP}} + \tau_{\mathrm{SWAP}})$.
Alternatively, we can use a whole row of ancillas between the two computational rows, and in $\left\lceil N / 2\right\rceil \tau_{\mathrm{CNOT}}$ prepare the cat state as we did without geometric constraints,
and again the CSWAP gates can be done in parallel.\footnote{Note that technically we should distinguish between different values for $\tau_{\mathrm{CSWAP}}$ depending on geometric constraints on the 2-qubit gates into which the CSWAP is decomposed, e.g., between when the control qubit is in the middle of the three on a line and when it is at one of the ends.}

\subsection{Diagonalization With Uncertainty}

\label{sec:diagonalization}

Given a collection of states
\(\{\ket{\phi_1},\ket{\phi_2}, \ldots, \ket{\phi_n}\}\), we are interested in determining
the minimum energy state in the subspace that they span. To do this, we
use our protocol described above to measure the matrix elements of the Hamiltonian and
overlap matrices (\eq{matrix_elements}),
and solve the generalized eigenvalue problem (\eq{generalized_eigenvalue_problem_1}).
However, because we perform only a finite number of measurements of each of matrix element, we have some level of statistical uncertainty. In this section, we shall lay out a simple Monte Carlo strategy to estimate the resulting uncertainty in the minimum eigenvalue of \eq{generalized_eigenvalue_problem_1}. We shall aim to provide a self-contained presentation for convenience, but we note that this approach is related to a long tradition of applying Monte Carlo methods to statistical problems, including the diagonalization of noisy matrices~\cite{Shinozuka1972-sj, Benaroya1992-ty, Soize2005-qf}.

We model the experimentally determined values of each matrix element using
a normal distribution.
In practice, the experimental measurements of the matrix elements are
individually described by draws from
Bernoulli random variables, but variational quantum algorithms typically work in the regime where the average of such measurements are well-approximated by a normal distribution~\cite{McClean2016-zj}.
In the context of an actual experiment, one could approximately determine the parameters of these distributions from the experimental measurement
record of the Hamiltonian and overlap
matrix elements.

For the purposes of the numerical experiments in this work, we determine the variance of the
Hamiltonian matrix element measurements using the upper bounds described in
Refs.~\citenum{McClean2016-zj}~and~\citenum{Rubin2018-ir}. Similarly, we observe
that our scheme for measuring the overlap matrix elements will have a variance that is
at most \(\frac{1}{m}\), where \(m\) is the number of measurements performed,
and we use this upper bound as an approximation to the true variance. We use
these approximations both in our simulation of the experimental measurement record
and in our subsequent protocol to determine the uncertainty in the ground state
energy.
Throughout this section, we use a notation which separates the intrinsic
component of the variance, which we denote by
\(\sigma^2\), from the scaling with the number of measurements, \(m\).

Experimentally, we only have access to estimates of
\(\mel{\phi_i}{\hat{H}}{\phi_j}\) and \(\braket{\phi_i}{\phi_j}\) from our
measurement record, which we denote by \(\tilde{h}_{ij}\) and
\(\tilde{s}_{ij}\). Taken together with our estimates of the variances,
\(\tilde{\sigma}^2_{H_{ij}}\) and \(\tilde{\sigma}^2_{S_{ij}}\), we can define the random variables
\begin{gather}
  \label{eq:H'_sample} \tilde{H}'_{ij} = \tilde{h}_{ij} +
\frac{\tilde{\sigma}_{H_{ij}}}{\sqrt{m_{H_{ij}}}}\mathcal{N}(0,1), \\ \tilde{S}_{ij}' = \tilde{s}_{ij} +
\frac{\tilde{\sigma}_{S_{ij}}}{\sqrt{m_{S_{ij}}}}\mathcal{N}(0,1).
  \label{eq:S'_sample}
\end{gather}
These distributions represent our uncertainty about the true value of the matrix
elements given the limited information provided by our experimental data.

To quantify the corresponding uncertainty in the ground state energy in the
NOVQE subspace, we use a Monte Carlo
sampling procedure. 
% maybe use a word other than ``integrate''?
We accomplish this by repeatedly drawing from the distributions
\(\tilde{H}_{ij}'\) and \(\tilde{S}_{ij}'\), and solving the resulting
generalized eigenvalue problems. However, it is possible that the noise in our matrix element measurements and
subsequent sampling destroys the positive semi-definite character of
the overlap matrix. To deal with this, we follow the canonical orthogonalization
procedure described in Ref.~\citenum{Szabo2012-ag}, discarding the
eigenvalues of the sampled overlap matrices that are less than some numerical
cutoff (and their associated eigenvectors). Each sampled pair of matrices yields
a sample from the unknown distribution over possible NOVQE ground state
energies. We then quantify our uncertainty in our estimate of this lowest
eigenvalue by calculating the sample variance of this distribution of possible energies, \(\sigma^2_{\mathrm{MC}}\).

It's important to note that this distribution is not Gaussian and that its mean is not an unbiased estimate of the ground state energy in the NOVQE subspace~\cite{Shinozuka1972-sj}. This is true for a number of reasons, but it can be seen, for example, by considering the fact that the usual second order correction to the energy is quadratic in the offdiagonal matrix elements. Therefore, even unbiased and normally distributed noise in the matrix elements leads to a bias in the estimated eigenvalues. Furthermore, the rate at which our estimate of the mean and variance of the distribution over possible NOVQE ground state energies convergences (with respect to the number of Monte Carlo samples) will vary based on the underlying distribution. The most meaningful consequences of this for our purposes is that convergence with respect to the number of Monte Carlo samples should be checked before being relied upon and that one should be cautious in using the standard error to generate error bars. As the number of measurements made increases and the amount of uncertainty diminishes these effects are naturally suppressed.

\subsubsection{Experiment Design Heuristic}
\label{sec:experiment_design}

In the previous section, we proposed a Monte Carlo scheme for estimating the
uncertainty in the NOVQE ground state energy caused by a finite number of
measurements of the individual matrix elements. By repeatedly sampling from \(\tilde{H}_{ij}'\) and \(\tilde{S}_{ij}'\) and solving the resulting generalized eigenvalue problems, we obtained a distribution over NOVQE ground state energies with some mean \(\mu_{\mathrm{MC}}\) and standard
deviation \(\sigma_{\mathrm{MC}}\). 
Here we build on this proposal to determine the relative impact of performing additional measurements. Ultimately, our goal is to create a reasonable heuristic for adaptively scheduling measurements to most efficiently use a limited amount of device time. 

We determine the impact of additional
measurements of the matrix elements on the uncertainty
in the ground state energy by calculating the derivatives of the sample standard
deviation, \(\sigma_{\mathrm{MC}}\), with respect to the number of measurements
performed, \(m_{H_{ij}}\) and \(m_{S_{ij}}\). The resultant quantities,
\(\frac{d\sigma_{\mathrm{MC}}}{d m_{H_{ij}}}\) and \(\frac{d\sigma_{\mathrm{MC}}}{d m_{S_{ij}}}\),
estimate how much we expect the sample deviation to shrink if we perform
additional measurements of \(H_{ij}\) or \(S_{ij}\). Note that we take these
derivatives only with respect to \(m_{H_{ij}}\) and \(m_{S_{ij}}\) in the Monte Carlo sampling procedure of
\eq{H'_sample} and \eq{S'_sample}, not the original measurements on
the device. Therefore, no additional quantum resources are required. We use the
TensorFlow software package to perform the Monte Carlo sampling of \(\tilde{H}_{ij}'\)
and \(\tilde{S}_{ij}'\), calculate of the ground state energies, and estimate \(\sigma_{\mathrm{MC}}\)~\cite{Abadi2016-pj}. This enables
us to evaluate the analytical expressions for each of
\(\frac{d\sigma_{\mathrm{MC}}}{d m_{H_{ij}}}\) and \(\frac{d\sigma_{\mathrm{MC}}}{d m_{S_{ij}}}\)
(for a fixed set of samples drawn from \(\tilde{H}_{ij}'\) and \(\tilde{S}_{ij}'\)) without
explicitly deriving the equations.

To optimally allocate our experimental measurements, we begin by
performing a small number of measurements of each matrix element. We then
estimate the derivatives \(\frac{d\sigma_{\mathrm{MC}}}{d m_{H_{ij}}}\) and
\(\frac{d\sigma_{\mathrm{MC}}}{d m_{S_{ij}}}\). Using these estimates, we simply
choose to perform additional measurements on the matrix element whose
corresponding derivative is the most negative. In practice, we perform these measurements
in small batches so that the time taken by the classical processing of the
measurement results is small compared to the time performing the measurements.
By repeating this process for many steps, until we either achieve the
desired accuracy or exhaust a pre-defined measurement budget, we aim to approximately optimize allocation of measurements between the different
terms.

\subsection{Implementation}

The tools presented above are applicable for use with
a variety of different ans\"atze, and subject only to the constraint that the
circuits act on a common reference state and conserve fermionic
excitation number in order to benefit from the efficient implementation of the matrix element measurements.
For our numerical experiments, we shall focus on a particular class of
wavefunctions known as $\textbf{\emph{k}}$-fold products of \textbf{u}nitary
\textbf{p}aired \textbf{c}oupled \textbf{c}luster with \textbf{g}eneralized
\textbf{s}ingle and \textbf{d}ouble excitations~\cite{Lee2019-ne}
($k$-UpCCGSD).
These wavefunctions have the appealing properties that 
1) % chktex 10
the required circuit depth scales only linearly in the size of the system, and
2) % chktex 10
they can be systematically improved by increasing the refinement parameter
\(k\). We briefly review this ansatz below and then describe in more detail the
implementation details of our numerical
experiments.

\subsubsection{The $k$-UpCCGSD Ansatz}

The essential idea behind the $k$-UpCCGSD ansatz is to act on a reference state,
Hartree-Fock in the case of this paper, with a product of \(k\) elementary
blocks. Each block is an independently parametrized approximation to a unitary
coupled cluster circuit generated by a sparse cluster operator containing only
single and paired double excitations~\cite{Stein2014-hn, Kowalski2018-lu}. To this end, the
wavefunction (before the Trotter approximation involved in compiling the
circuits) is defined as follows.
\begin{equation} |\psi\rangle = \prod_{x = 1}^{k} \left( e^{\hat{T}^{(x)}-
{\hat{T}^{(x)}}^\dagger} \right)|\phi_0\rangle,
\label{eq:kup_overview}
\end{equation}
where each cluster operator
\begin{equation} \hat{T} = \sum_{ia} {t_{i i}^{a a}} \hat{a}^\dagger_{a\alpha}
\hat{a}^\dagger_{a\beta} \hat{a}_{i\beta} \hat{a}_{i\alpha} + t_i^a(
\hat{a}^\dagger_{a\alpha} \hat{a}_{i\alpha} + \hat{a}^\dagger_{a\beta}
\hat{a}_{i\beta} ).
\label{eq:Tpccd}
\end{equation}
possesses an independent collection of variational parameters.
(We omit the \((x)\) superscript for simplicity and use Latin and Greek letters for
spatial and spin indices respectively.)

In contrast with the standard unitary coupled cluster single and doubles
(UCCSD), $k$-UpCCGSD only includes doubles excitations which collectively move
a pair of electrons from one spatial orbital to another. The resulting loss of flexibility is ameliorated by the use of
generalized excitations that do not distinguish between occupied and unoccupied
orbitals~\cite{Nakatsuji2000-fd, Nooijen2000-ji}, and the k-fold repetition of the
elementary circuit block. As a result, the number of free parameters in the
ansatz scales as \(O(kN^2)\). We make use of the generalized swap networks of Ref.~\citenum{OGorman2019-qz} to implement a single Trotter step approximation to the $k$-UpCCGSD ansatz with the open source
Cirq and OpenFermion-Cirq libraries~\cite{The_Cirq_Developers2019-ke}.

The circuits consist of the following gates:
\begin{itemize}
\item 
$\mathrm{FSIM}_2(w_0, w_1) = \exp(i H)$ for \\
$H = 
\left(
    w_0 \ket{10}\bra{01} + \mathrm{h.c.}
\right) +
w_1 \ket{11}\bra{11}$,
\item
$\mathrm{FSWAP} = \mathrm{SWAP} \cdot \mathrm{CZ}$, and
\item
$\mathrm{FSIM}_4(w) =
\exp(i H)$ for  \\
$H = w \ket{0011}\bra{1100} + \mathrm{h.c.}$
\end{itemize}
Because each FSWAP immediately follows an $\mathrm{FSIM}_2$, we can compile them together to get an effective duration $\tau_2$.
Let $\tau_4$ be the effective duration of $\mathrm{FSIM}_4$.
The overall depth then is $k N (\tau_2 + \tau_4 /2)$.
There are $\binom{N}{2}$ pairs of $\mathrm{FSIM}_2$ and FSWAP gates, and $\binom{N/2}{2}$ $\mathrm{FSIM}_4$ gates.
This is simply an upper bound; the depth may be compressed further by combining the compilation of each $\mathrm{FSIM}_4$ with the immediately following 2-qubit gates.
  
\subsubsection{Computational Details}
\label{sec:computational_details}

The quantum chemical
calculations of the full configuration interaction (FCI) ground states and Hartree-Fock (HF) reference wavefunctions were performed using the open
source packages Psi4 and OpenFermion~\cite{Parrish2017-yw, McClean2017-cs}. 
We optimized the ground state energy in the NOVQE subspace by varying the
parameters of the most recently added ansatz wavefunction, diagonalizing the Hamiltonian
and overlap matrices at each step. Inspired by recent proposals for adaptive ansatz construction~\cite{Grimsley2019-qa, Tang2019-kr, Ryabinkin2020-fz}, each k-UpCCGSD wavefunction was grown iteratively by adding a single UpCCGSD block at a time, as described in more detail below. We performed this optimization using the Scipy implementation of the quasi-Newton limited-memory
BFGS (L-BFGS-B) method~\cite{Jones_undated-fr, Zhu1997-ia}, treating the ground state energy in the NOVQE subspace as the objective function. We calculated the gradient at each step using a finite difference method with a step size of \(\delta=10^{-6}\). Each circuit was optimized using up to 2000 gradient evaluations. 

In order to escape local minima, we repeatedly applied random kicks to the variational parameters. After each 500 gradient evaluations we compared the current value of the objective function to the best observed value and reset the parameters if appropriate. Subsequently, we added random values drawn from the a normal distribution with zero mean and variance \(\sigma^2=1\) (after 500 steps), \(\sigma^2=10^{-1}\) (after 1000 steps), or \(\sigma^2=10^{-2}\) (after 1500 steps). The best observed value of the energy across this whole procedure is the one we ultimately report. We randomly initialized the parameters of the \(k=1\) circuits by drawing from a normal distribution with mean 0 and variance \(\sigma^2 = 10^{-6}\). Parameters for circuits with higher values of \(k\) were initialized by taking the parameters from an optimized circuit with \(k-1\) UpCCGSD blocks and appending a new block with random variational parameters drawn from the same distribution, \(\mathcal{N}(0, 10^{-6})\).

\section{Results}

H$_4$ is often used as a small
testbed for single-reference coupled-cluster
methods~\cite{Paldus1993,Mahapatra1999,Kowalski2000,Jankowski1980,Evangelista2006}.
We shall focus on the square
(D$_\text{4h}$) geometry here. The system exhibits two exactly degenerate determinants at the
D$_\text{4h}$ geometry, leading to a mix of strong and weak correlation effects.
Another important class of chemical systems to investigate is
hydrocarbons. In this work, we shall study a simple hydrocarbon,
hexatriene (C$_6$H$_8$). The interesting aspect of this molecule is that the
torsional PES of a double bond leads to a strong correlation problem. At $\theta
= 90^\circ$, it exhibits two exactly degenerate determinants and therefore it is
strongly correlated. To form the active space, we include the entire set of $\pi$ electrons in the system
along with both $\Pi$ and $\Pi^*$ orbitals. The resulting active space is then
(6e, 6o), and this also possesses a good mixture of weak and strong correlation.

In the following subsections, we present the results of two types of experiments related to our proposed NOVQE approach on these
chemical systems and discuss the potential utility of our approach for more
general chemical problems. With the first class of experiments, we focus on
understanding how effectively the ground state can be represented by a linear
combination of parametrized wavefunctions, optimized using the gradient-based
approach we described above. We vary both the complexity of the individual
ansatz wavefunctions by adjusting the number of circuit blocks (\(k\)) in the
$k$-UpCCGSD ansatz and the number of states \((M)\) in the NOVQE subspace. For these calculations, we neglect the challenges posed by a finite number of measurements and the impact of circuit noise. In our second set of numerical experiments, we explore the extent to which our proposal for an adaptive measurement scheme is successful in reducing the number of circuit repetitions required to resolve the NOVQE ground state energy to a fixed precision.

\subsection{NOVQE Ground State Energies}

\subsubsection{A Hydrogen Complex, \ce{H4}}

\begin{figure}[t] \centering
\includegraphics[width=.5\textwidth]{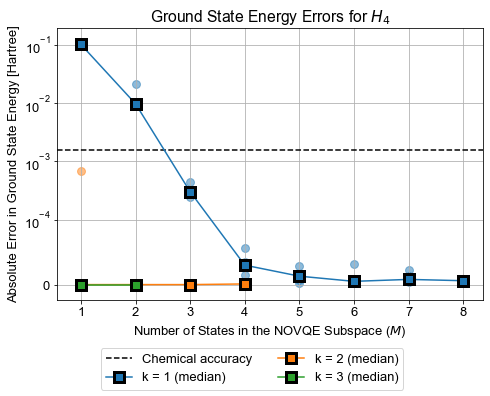}
  \caption{Difference between NOVQE energies and FCI energies for the ground
    state of H$_4$ for a variety of k-UpCCGSD ans\"atze and sizes of the NOVQE subspace (M).
    The
NOVQE energy is optimized by varying the parameters of the most recently added
state to minimize the ground state energy in the subspace. For each value of M and k we plot five independent calculations as separate points and show the median values as squares connected by lines. The dotted
horizontal line indicates 1 kcal/mol \(\approx\) 1.59 millihartree, a commonly
accepted value for ``chemical accuracy''. As more states are added to the NOVQE
subspace, the error in the ground state energy declines substantially for the
\(k=1\) version of $k$-UpCCGSD. For larger values of k, a single state (equivalent to a regular VQE procedure) is sufficient to capture the ground state to a high precision.}
  \label{fig:h4_abs}
\end{figure}

\fig{h4_abs} presents data on the
application of NOVQE to the square geometry of H\(_4\) with fixed bond distance
$R_{\text{H-H}} = 1.23\:\mbox{\normalfont\AA}$ in a minimal STO-3G basis set, an
\(N = 8\) qubit problem. We consider the performance of the $k$-UpCCGSD ansatz for
\(k=1\) to \(k=3\) with \(M=1\) up to \(M=8\) states in the NOVQE
subspace, noting that \(M=1\) is equivalent to the regular VQE procedure. 
%For
%the sake of comparison we also show the performance of VQE with the trotterized
%UCCSD ansatz. 
For each value of k and M we perform five independent calculations and consider the error in the median ground state
energy found by the optimization procedure as a proxy for ansatz's ability to describe the ground state.

Focusing first on understanding the behavior of the wavefunctions in the
context of the standard VQE approach (\(M=1\)), we note that for \(k \geq 2\) the k-UpCCGSD ansatz is essentially exact for this problem. Looking more closely at the data for \(k=2\), \(M=1\) in \fig{h4_abs}, one can see that one of the five calculations failed to find the global optimum (the pale orange point). In general, we found that the optimization of this ansatz was challenging. We expect these challenges to become more severe with increasing system size, and when the stochastic nature of the quantum measurements are taken into account.

In the case of \(k=1\) we observe that we can systematically
improve the accuracy of the estimated ground state energy by increasing the
number of states included in the NOVQE subspace (\(M\)). Given \(M = 3\) independent copies, even this relatively simple ansatz is able to represent that ground state almost exactly. 
This supports our thesis that a collection of
ansatz states which are individually not capable of targetting a desired state may be
fruitfully combined to yield a sufficiently powerful logical ansatz. However, the measurements of the offdiagonal matrix elements for NOVQE require slightly more than twice the gate count necessary for the measurements of  individual ansatz states in the regular VQE formalism. For this particular system, it may therefore be more effective to use a single \(k=2\) ansatz than multiple \(k=1\) circuits.

\subsubsection{Hexatriene}

\begin{figure*}[t] \centering
\includegraphics[width=\textwidth]{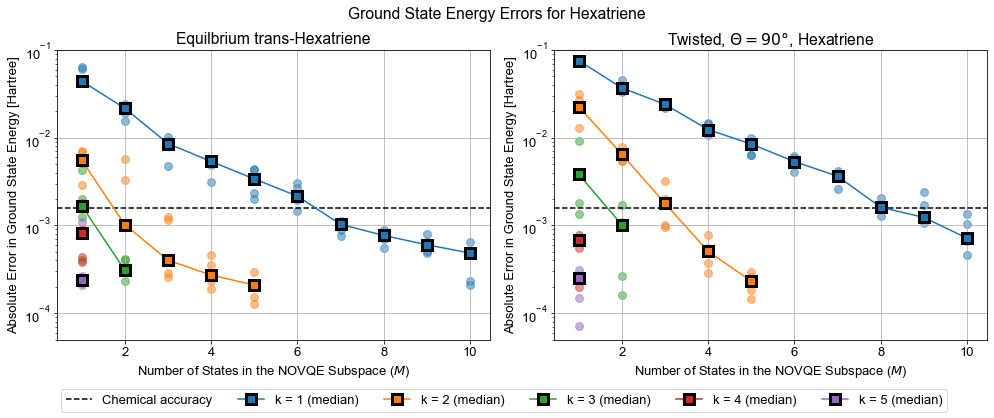}
  \caption{Difference between NOVQE energies and FCI energies for the ground
states of the equilibrium configuration of trans-Hexatriene and a \(90^\circ\) twisted configuration for a variety of k-UpCCGSD ans\"atze and sizes of the NOVQE subspace (M).
The
NOVQE energy is optimized by varying the parameters of the most recently added
state to minimize the ground state energy in the subspace.
For each value of M and k we plot five independent calculations as separate points and show the median values as squares connected by lines. The dotted
horizontal line indicates 1 kcal/mol \(\approx\) 1.59 millihartree, a commonly
accepted value for ``chemical accuracy''. The flexibility of the NOVQE wavefunction may be increased both by adding more states to the NOVQE subspace (M), or more parametrized blocks to each individual circuit (k). In either case, the error is driven below the threshold for chemical accuracy.}
  \label{fig:hexa_two_abs}
\end{figure*}

Here we present our results for the ground state energy of two molecular configurations of
Hexatriene (C\(_6\)H\(_8\)) in an STO-3G basis with an active space of \(6\)
electrons in \(6\) $\pi$ orbitals (\(N = 12\) qubits).
Here, due to the system's increased complexity, we consider circuits with up to $k=5$ UpCCGSD blocks and subspace sizes as large as \(M = 10\). In \fig{hexa_two_abs} we
show the calculations for an equilibrium geometry (the trans isomer, obtained by performing
geometry optimization using density functional theory) and a configuration with
a \(90^\circ\) twist on the central Carbon-Carbon double bond respectively. We
provide the geometries for these two configurations in \app{hexatriene_geometries},  \tab{trans_hexatriene_coords}
and \tab{90_deg_hexatriene_coords}.

Once again we notice that increasing the circuit complexity by
taking larger values of \(k\) provides a substantial benefit, driving the estimated ground state energy well below the threshold for chemical accuracy without resorting to the multiple states of the NOVQE formalism. Likewise, as the number of NOVQE states increases, the NOVQE ground state energy reaches chemical accuracy even with the most limited ansatz. For Hexatriene we see that multiple \(k=1\) states are able to achieve a performance on par with a single \(k=4\) state. The NOVQE procedure for the \(k=1\) states requires almost a factor of four less circuit depth and half as many quantum gates as performing VQE with the \(k=4\) state.

Interestingly, for the \(k=1\) case in both configurations, and the \(k=2\) case in the twisted configuration, \fig{hexa_two_abs} shows the error in the ground state energy decreasing exponentially as a function of \(M\). We contrast this with the observation in classical 
non-orthogonal
electronic structure calculations, where a small number of determinants are often sufficient to capture most of the wavefunction, but a long tail of dynamic correlation can result in a slow convergence to the true ground state
as determinants are added to the variational space~\cite{Koch1993-mo, Malmqvist1986-nz, Thom2009-fp, Jimenez-Hoyos2013-wb, McClean2015-es, Sundstrom2014-rb, Landinez_Borda2019-bj, lee2019auxiliary}. 
The classical intractability of
calculating matrix elements between different coupled cluster wavefunctions
means that relatively little work has been done on the representational power of
wavefunctions like those used in NOVQE. We speculate here that the increased flexibility of the individual ansatz states may be allowing for a good representation of the ground state to be achieved before entering a regime of slow convergence. 
This is in contrast with another class of quantum non-orthogonal methods which, by virtue of building their basis states by time-evolving a set of reference wavefunctions, demonstrate exponential convergence by construction~\cite{Parrish2019-hf, Stair2019-sc}.
In the future, it would be interesting to determine whether the relatively quick convergence with respect to \(M\) we observe here breaks down for more complicated systems when a purely variational approach to constructing the basis states is taken.

\subsection{NOVQE Matrix Element Measurements}

\begin{figure*}[t]
  \centering
  \includegraphics[width=\textwidth]{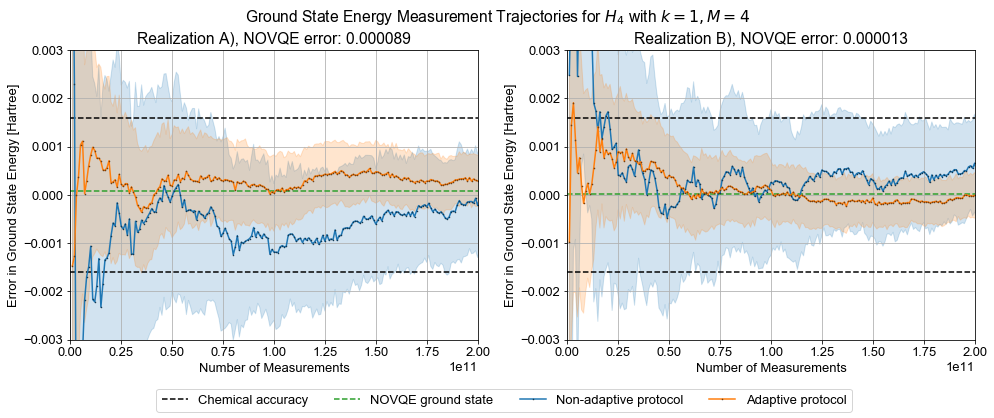}
  \caption{Comparison of the ability of the adaptive and non-adaptive schemes for
    scheduling measurements to resolve the ground state energy of H\(_4\) in 
    two different NOVQE subspaces of \(M=4\) optimized \(k=1\) $k$-UpCCGSD states. The evolution of
    the estimated ground state energies is plotted in solid lines together with
    \(2 \sigma\) error
  bars indicated by the shaded regions. The actual energies of the ground states in
the NOVQE subspaces are indicated with dashed green lines. Panels A and B show two different typical realizations of the measurement record as the number of measurements increases. In both cases, the adaptive
protocol converges significantly more quickly than the non-adaptive one. Note that the variance of the experimental measurements are approximated using upper bounds and that the true numbers required for both the adaptive and non-adaptive schemes are likely to be lower~\cite{McClean2016-zj, Huggins2019-ej}.}
  \label{fig:h4_k2n4_two}
\end{figure*}

In the previous subsection we presented data on the
performance of NOVQE in the absence of noise during the circuit
execution and measurement process. Now we consider the effects of statistical
noise during measurement. Specifically, we determine how many
circuit repetitions are necessary to evaluate the ground state energy within a
target precision for a subspace defined by a fixed set of NOVQE states. For
simplicity, we do not combine this analysis with an investigation of the
optimization procedure. Instead, we take the optimized circuit parameters for a
collection of \(M\) NOVQE states and compare the effectiveness of the adaptive protocol
we described in \sec{experiment_design} to a simpler alternative for determining
the ground state energy in the subspace spanned by the optimized states, which we
shall explain below.

The simpler protocol, which we shall refer to as non-adaptive, consists of
measuring each matrix element of the Hamiltonian and overlap matrices the
same number of times. For the adaptive protocol, we repeatedly use the procedure
described in \sec{experiment_design} to select a
particular matrix element and perform measurements in batches of \(\approx 10^5\) circuit repetitions. For the
purpose of this comparison, we treat a `measurement' of a particular Hamiltonian or
overlap matrix element as a draw from a Gaussian random variable whose mean is
the true value of the matrix element and whose variance is set by the upper
bound described in Ref.~\citenum{McClean2016-zj}, scaled by the number of
measurements performed. Note that in a real experiment, or a finer-grained simulation, the
Hamiltonian has to be decomposed into groups of terms that can be
simultaneously measured, and one could apply an adaptive scheme like the one we
propose to schedule measurements between these groups as well. For both kinds of numerical experiments we calculate a
\(2\sigma\) error bar using a bootstrapping sample size of 200 using the
techniques of \sec{diagonalization}.

\subsubsection{A Hydrogen Complex, \ce{H4}}

In \fig{h4_k2n4_two} we plot the actual trajectories of the estimates for the
ground state energies, together with their error bars for both the adaptive and
non-adaptive approaches to measurement. We show two realizations of this
numerical experiment applied to an NOVQE simulation of \ce{H4} with \(M=4\) 1-UpCCGSD states. In both cases, we see that the adaptive protocol converges
more quickly towards the NOVQE ground state energy than the non-adaptive one. We find that the data qualitatively supports the assumption that the variance in
the ground state energy estimate settles into an asymptotic regime where its
behavior is well described by the relationship
\begin{equation}
  \sigma^2(N) \approx \frac{\kappa}{N},
  \label{eq:variance_scaling}
\end{equation}
  where \(N\) indicates the number of measurements performed and
\(\kappa\) is some constant. For these particular realizations, we find \(\kappa\) to be
approximately 
\( 5.3 \cdot 10^4~E_h^2\) and \( 5.5 \cdot 10^4~E_h^2\) for the non-adaptive scheme in panels A) and B), and approximately \( 1.4 \cdot 10^4~E_h^2\) and \( 9.7 \cdot 10^3~E_h^2\) for the adaptive
ones. Using the same upper bounds to calculate the variance for a regular VQE
calculation performed on the same system would yield \(\kappa \approx 28~E_h  ^2\). 

Therefore, for these
applications to \ce{H4}, our scheme for iterative measurement achieves a modest reduction in variance. When targeting a fixed accuracy this would translate into a few-fold (\(\approx 3.7\) for realization A and \(\approx 5.7\) for realization B) savings in measurement time. 
Unfortunately, this cost is orders of magnitude larger than that required for energy measurement in an ordinary VQE approach. In order for NOVQE, or other forms of quantum non-orthogonal methods to be made practically useful, this increased measurement time will have to be accounted for and minimized.

\begin{figure*}[t]
  \centering
  \includegraphics[width=\textwidth]{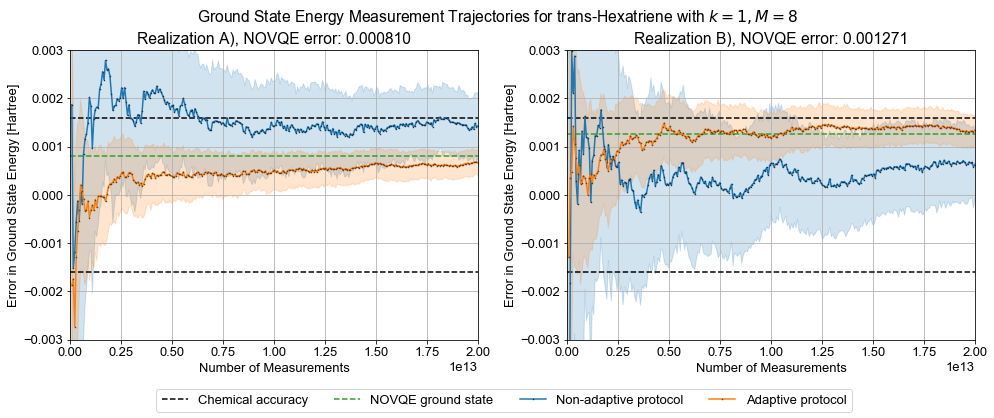}
  \caption{Comparison of the ability of the adaptive and non-adaptive schemes for
    scheduling measurements to resolve the ground state energy of trans-Hexatriene in 
    two different NOVQE subspaces of \(M=8\) optimized \(k=1\) $k$-UpCCGSD states. The evolution of
    the estimated ground state energies is plotted in solid lines together with
    \(2 \sigma\) error
  bars indicated by the shaded regions. The actual energies of the ground states in
the NOVQE subspaces are indicated with dashed green lines. Panels A and B show two different typical realizations of the measurement record as the number of measurements increases. In both cases, the adaptive
protocol converges significantly more quickly than the non-adaptive one. Note that the variance of the experimental measurements are approximated using upper bounds and that the true numbers required for both the adaptive and non-adaptive schemes are likely to be lower~\cite{McClean2016-zj, Huggins2019-ej}.}
  \label{fig:hexa_k2n16}
\end{figure*}

\subsubsection{Hexatriene}

As in our analysis of H\(_4\), we compare the proposed adaptive approach to
distributing measurements between the elements of the Hamiltonian and overlap
matrices with a non-adaptive one. We do so by choosing collections of optimized NOVQE
states and applying both methods to determine the ground state energy in the
resulting subspaces. In this case we choose to use \(M=8\) states, each of which
is generated by a \(k=1\) $k$-UpCCGSD circuit, and focus on the equilibrium
configuration of trans-Hexatriene. Examining the two realizations of this experiment plotted in \fig{hexa_k2n16}, we see immediately
that the increased difficulty of this problem compared to H\(_4\) is reflected in the much larger gaps between the FCI ground states and the ground states
in the NOVQE subspaces, as well in the larger numbers of measurements required for
convergence.

\fig{hexa_k2n16} shows the same substantial difference between the performances
of the adaptive and non-adaptive approaches that was observed for \ce{H4}. In panel B, we
see that the true ground state of the subspace lies outside of the error bars
for the non-adaptive scheme during small portions of the measurement procedure.
This is a manifestation of the phenomenon mentioned in \sec{diagonalization}, where using an insufficient number of Monte Carlo samples may result in misestimating the magnitude of the uncertainty in the ground state energy.
We note that the
adaptive scheme moves quickly to a regime where the
uncertainty estimates are reliable even with a small number of samples. We
once again observe that the variance qualitatively converges with the
expected long-time \(\frac{1}{N}\) behavior of \eq{variance_scaling} for most of
the numerical experiment. Therefore, we can determine \(\kappa\), the `intrinsic
variance' defined in \eq{variance_scaling}, of each method and compare their statistical efficiencies.

For the non-adaptive scheme we observe
\(\kappa \approx 2.4 \cdot 10^6\) and \(\kappa \approx 2.9 \cdot 10^6\) for panels A and B, while for the adaptive scheme we see \(\kappa \approx
3.7 \cdot 10^5\) and \(\kappa \approx 6.5 \cdot 10^5\). The reference value for a regular VQE calculation is  \(\kappa \approx 1.6 \cdot 10^2\), determined using the same bounds assumed
throughout this comparison~\cite{McClean2016-zj,Rubin2018-ir}. Comparing with the simpler H\(_4\) example, we see that
the adaptive scheme for measuring the NOVQE ground state energy of Hexatriene
results in a slightly larger gain when compared to the non-adaptive scheme, but still
falls short of the goal of reducing the number of measurements to an experimentally plausible number. One
promising avenue to further reducing this cost is the adaptation of recently proposed strategies for
measurement in the context of regular VQE to NOVQE~\cite{Huggins2019-ej}. These
strategies have been shown to reduce the number of circuit executions by orders
of magnitude when compared with the bounds used to derive the number of measurements in this work. Further study is required in order to determine if one can alter the optimization process of the NOVQE states themselves or their coefficients in order to achieve an additional reduction in the measurement cost.

\section{Discussion and Outlook}

We have introduced an extension to the
variational quantum eigensolver that calls for the ground state energy to be
approximated by solving a generalized eigenvalue problem in a subspace that is spanned
by a linear combination of \(M\) parametrized quantum wavefunctions. The resulting
logical wavefunction ansatz is a linear combination of all \(M\) states in the
subspace, but its properties can be determined by only pairwise measurements of the Hamiltonian and
overlap matrices. Therefore, it is possible to increase the flexibility of the
ansatz without requiring additional coherent quantum resources. By analogy with
the non-orthogonal configuration interaction method of classical quantum
chemistry~\cite{Malmqvist1986-nz, Thom2009-fp, Sundstrom2014-rb}, we call our approach the non-orthogonal variational quantum
eigensolver, NOVQE.\@

Our proposal necessitates off-diagonal measurements of the Hamiltonian and
overlap matrices. We perform these using a modified Hadamard test. Naively, this would
require us to implement controlled versions of the quantum circuits for state
preparation. To avoid this cost, we demanded that the state
preparation circuits all act on a common reference state and preserve fermionic
excitation number. This allowed us to avoid the need to add controls to the
ansatz circuits, by instead performing controlled swap operations between two
copies of the system register, a cost that scales linearly and modestly with the
system size.

To determine the ground state energy in the subspace, our approach
requires that we measure all \(M^2\) elements of the Hamiltonian and overlap matrices
in the NOVQE subspace. We presented a statistical strategy for estimating the
uncertainty in the resultant ground state energy estimate for a given
uncertainty in the matrix elements. We also pointed out how the machinery that
generates these estimates can be leveraged in a Monte Carlo sampling process
to determine which matrix element
should be chosen for additional measurements to optimally reduce the
uncertainty. We proposed an iterative approach, in which small batches of
measurements are repeatedly performed according to this Monte Carlo prescription, to
minimize the overall number of circuit repetitions required by our NOVQE method.

We demonstrated an implementation of our approach using a collection of
$k$-UpCCGSD wavefunctions to approximate the ground state of two model
strongly-correlated systems, a square geometry of H\(_4\) and the \(\pi\)-space
of Hexatriene in two configurations. Growing the NOVQE subspace by adding and
optimizing one state at a time,
we showed how a collection of ans\"atze which individually struggle to represent the ground state can be fruitfully combined
combined to form a more powerful logical ansatz. In our numerical experiments we observed that the marginal utility of adding additional states to the NOVQE subspace remained large, even as the size of the space increased. This is in contrast with the commonly noted behavior of classical non-orthogonal methods, which generate a collection of non-orthogonal Slater determinants and diagonalize in the resulting subspace~\cite{Koch1993-mo, Malmqvist1986-nz, Thom2009-fp, Jimenez-Hoyos2013-wb, McClean2015-es, Sundstrom2014-rb, Landinez_Borda2019-bj}. These approaches eventually enter a regime where convergence slows down significantly as states are added to the subspace. This suggests that the there is a benefit in NOVQE's ability to make use of wavefunctions more sophisticated than the Slater determinants available to classical non-orthogonal methods, allowing for a balance between the number of distinct wavefunctions and their flexibility.

To characterize our proposal for adaptively scheduling measurements to
minimize the number of circuit repetitions required by our approach, we
focused on quantifying the number of measurements required to approximate the
ground state energy in a fixed NOVQE subspace. For the purposes of this
investigation we approximated the variance of the individual matrix element measurements using
the bounds described in Refs.~\citenum{McClean2016-zj}~and~\citenum{Rubin2018-ir}. For both our square
H\(_4\) and our equilibrium configuration of trans-Hexatriene, we optimized collections of NOVQE states and froze their parameters.
We then applied our adaptive approach for
scheduling measurements and compared it to a simpler non-adaptive scheme, in which
each matrix element was measured the same number of times. We found that our
adaptive approach used somewhat fewer measurements than a simpler non-adaptive strategy, but dramatically more than it would take to measure the energy in the standard VQE formalism. It would be worthwhile to understand whether similar challenges appear for other proposed quantum non-orthogonal methods~\cite{Motta2020-in, Parrish2019-ev, Kyriienko2020-ez, Parrish2019-hf, Stair2019-sc}.

We can imagine several routes towards ameliorating this difficulty and
developing NOVQE further. First, having states
that are nearly linearly dependent in the NOVQE subspace can dramatically
increase the cost of measurement. Developing an optimization strategy for the individual states, or their coefficients, that
regularizes this behavior away would be useful. Related to this is the possibility of extending the tools for measuring
analytical gradients of parametrized quantum circuits to work with the NOVQE
formalism. Another avenue for future work would be the development of
good initialization strategies for NOVQE, potentially using reference states
derived from a classical NOCI
calculation. Finally, recent work has shown that a measurement strategy based on
factorizations of the two-electron integral tensor can dramatically
reduce the cost of the standard VQE approach, lowering the number of separately measured terms from \(O(N^4)\) to \(O(N)\) ~\cite{Huggins2019-ej}. The resulting cost reduction is especially large when compared to the
type of bounds used throughout this paper~\cite{McClean2016-zj,Rubin2018-ir}.
Adapting this approach for use with NOVQE is likely to offer a significant improvement.

Beyond these modifications to the NOVQE approach outlined in this paper, it is
also conceivable that the tools we have presented might be usefully employed in
other ways. For example, we have focused here on the variational optimization of a logical ansatz
that is a superposition of individual parametrized wavefunctions. An alternative
is to take inspiration from Ref.~\citenum{Motta2020-in}, and from the classical
NOCI method~\cite{Malmqvist1986-nz, Thom2009-fp, sundstrom2014non}, and optimize
the individual wavefunctions separately, solving the generalized eigenvalue problem only once with the
final collection of states. In this vein, there are several recent proposals which form a non-orthogonal basis using a collection of time-evolved reference states~\cite{Parrish2019-hf, Stair2019-sc}. Another possible direction to pursue is the inclusion of one
or more states in the NOVQE subspace that can be classically optimized, only turning to the
use of more general parametrized quantum circuits to prepare small corrections to the classically tractable states. All of these ideas have the potential to benefit from the tools we have developed for efficiently performing the required matrix element measurements.

In summary, this work has presented a promising new extension to the VQE
formalism and highlighted both its advantages and its drawbacks.
We have also presented a strategy for compiling off-diagonal
matrix element measurements and promoted a general approach to Monte Carlo
estimation of uncertainty, which may be of independent interest. The circuit simulations of the $k$-UpCCGSD ansatz presented here add to the analyses of
Refs.~\citenum{Lee2019-ne},~\citenum{Grimsley2020-mq}, and~\citenum{Matsuzawa2020-ad}. We believe that the ability of our NOVQE
to trade off coherent quantum resources for additional measurements may prove to
be a useful tool in making use of NISQ-era quantum hardware for studying
challenging strongly correlated systems.

In the final stages of preparing this manuscript two works were posted
which independently developed approaches using the matrix elements
between collections quantum states for other applications. One appears in the context
of variational quantum algorithms for solving linear systems of
equations~\cite{Huang2019-cn}, while the other proposes a strategy for approximating
the low energy subspace of a Hamiltonian in terms of time-evolved trial wavefunctions~\cite{Parrish2019-hf}.

\section{Acknowledgements}

This work
was supported by the U.S. Department of Energy, Office of Science, Office of Advanced Scientific Computing
Research, Quantum Algorithm Teams Program, under contract number DE-AC02-05CH11231. Additionally, this work was supported by a Quantum Algorithms Focused Award from Google LLC.\@ B.O. was supported by a NASA Space Technology Research Fellowship.

% \bibliography{paperpile,manually_added}{}
\input{main.bbl}

\newpage

\appendix

\section{Hexatriene Geometries}
\label{app:hexatriene_geometries}
The equilibrium geometry was obtained from the geometry optimization with $\omega$B97X-D~\cite{chai2008long} and cc-pVTZ~\cite{Dunning1989} using a development version of Q-Chem\cite{Shao2015}.
The \(90\degree\) twisted configuration was obtained by rotating the middle C-C double bond out-of-plane.

% chktex-file 37
% chktex-file 44
\begin{table}[h] \centering
  \begin{tabular}{c|r r r}
    Atom & X & Y & Z \\
    \hline
    C & ( 0.5987833, &  0.2969975, &  0.0000000) \\
    H & ( 0.6520887, &  1.3822812, &  0.0000000) \\
    C & (-0.5987843, & -0.2970141, &  0.0000000) \\
    H & (-0.6520904, & -1.3822967, &  0.0000000) \\
    C & (-1.8607210, &  0.4195548, &  0.0000000) \\
    H & (-1.8010551, &  1.5036080, &  0.0000000) \\
    C & (-3.0531867, & -0.1693136, &  0.0000000) \\
    H & (-3.9685470, &  0.4053361, &  0.0000000) \\
    H & (-3.1479810, & -1.2485605, &  0.0000000) \\
    C & ( 1.8607264, & -0.4195599, &  0.0000000) \\
    H & ( 1.8010777, & -1.5036141, &  0.0000000) \\
    C & ( 3.0531816, &  0.1693296, &  0.0000000) \\
    H & ( 3.9685551, & -0.4052992, &  0.0000000) \\
    H & ( 3.1479561, &  1.2485793, &  0.0000000) \\
  \end{tabular}
  \caption{The geometry of the equilibrium configuration of trans-Hexatriene.}
  \label{tab:trans_hexatriene_coords}
\end{table}
\begin{table}[b] \centering
  \begin{tabular}{c|r r r}
    Atom & X & Y & Z \\
    \hline
    C & ( 0.5987833, &  0.2969975, &  0.0000000) \\
    H & ( 1.3716346, & -0.0683717, &  0.6707370) \\
    C & (-0.5987843, & -0.2970141, &  0.0000000) \\
    H & (-1.3716354, &  0.0683544, &  0.6707361) \\
    C & (-0.9484080, & -1.4197297, & -0.8504282) \\
    H & (-0.1721763, & -1.7803215, & -1.5183873) \\
    C & (-2.1390983, & -2.0121775, & -0.8520831) \\
    H & (-2.3554088, & -2.8468591, & -1.5037144) \\
    H & (-2.9353514, & -1.6772360, & -0.1982062) \\
    C & ( 0.9484189, &  1.4197134, & -0.8504230) \\
    H & ( 0.1721980, &  1.7803171, & -1.5183881) \\
    C & ( 2.1391167, &  2.0121462, & -0.8520613) \\
    H & ( 2.3554502, &  2.8468291, & -1.5036834) \\
    H & ( 2.9353585, &  1.6771903, & -0.1981764) \\
  \end{tabular}
  \caption{The geometry of the \(90\degree\) twisted configuration of Hexatriene.}
  \label{tab:90_deg_hexatriene_coords}
\end{table}
% chktex-file 17
\end{document}